\documentclass[11pt,a4paper, notitlepage]{article}
\usepackage[utf8]{inputenc}
\usepackage{authblk}
\usepackage{graphicx}

\title{Unidirectional guided resonances in anisotropic waveguides}
\author[1]{Samyobrata Mukherjee}
\author[1]{Jordi Gomis-Bresco}
\author[1,2,*]{David Artigas}
\author[1,2]{Lluis Torner}

\affil[1]{ICFO-Institut de Ciències Fotòniques, The Barcelona Institute of Science and Technology, 08860 Castelldefels (Barcelona), Spain}
\affil[2]{Department of Signal Theory and Communications, Universitat Politècnica de Catalunya, 08034 Barcelona, Spain}
\affil[*]{david.artigas@icfo.es}
\date{}

\begin{document}

\maketitle

\begin{abstract}
We show that anisotropic planar anti-guiding waveguide structures with two radiation channels towards the surrounding cladding materials can support unidirectional guided resonances (UGRs), where radiation is cancelled in one of the radiation channels and redirected into the other. Their formation is subtle as it requires breaking the so-called polar {\it anisotropy-symmetry\/} of the structures. Then, UGRs appear at specific wavelengths and light propagation directions, are robust, and are characterised by phase singularities in the  channel in which radiation is cancelled. The mechanism we describe allows for ready selection of the radiation direction, as well as tuning of the wavelength and the propagation angle at which UGRs occur. 
\end{abstract}

Unidirectional guided resonances (UGRs) are unbounded states of photonic structures where light is radiated from a waveguide via only one radiation channel even though multiple channels may be available. Their existence has been recently put forward and experimentally observed in photonic crystal structures \cite{Yin2020}. By and large, control of the proportion of radiation escaping via each radiation channel is desirable for various applications, such as photonic crystal surface emitting lasers \cite{Hirose2014}, vertical grating couplers \cite{Taillaert2004, Vermeulen2010} or light detection and ranging devices \cite{Yaacobi2014}, to name a few. Several schemes have been proposed to obtain unidirectional radiation by directing all radiation into a single  channel, for example using stacked reflectors \cite{Roncone1993,Kim2006}, interference from radiating antennae \cite{Wade2015}, or asymmetric photonic crystal structures \cite{Wang2013, Ota2015, Zhou2016}. However, the aim of perfectly unidirectional radiation was only achieved using the concept of topologically enabled radiation cancellation that is typical of bound states in the continuum (BICs) \cite{Yin2020}. BICs are states that remain radiationless even though they exist embedded in the continuum part of the spectrum \cite{Neuman1929, Stillinger1975}. They have been a topic of intense recent study \cite{Marinica2008, Bulgakov2008, Plotnik2011, Corrielli2013, Hsu2013,  Gomis-Bresco2017, Monticone2017, Minkov2018, Azzam2018, Fan2019, Liang2020}, which has led to several potential applications \cite{Kodigala2017,Romano2018, Carletti2018, Hayran2021}. Photonic BICs arise due to symmetry protection or destructive interference via parameter tuning \cite{Hsu2016}. BICs are zeroes of radiation by definition and therefore correspond to polarisation or phase singularities and may exhibit topological properties \cite{Zhen2014, Bulgakov2017, Doeleman2018, Zhang2018, Mukherjee2018, Jin2019}. The insight into the mechanism of radiation cancellation gained from BICs has been harnessed in structures with multiple radiation channels to create directional resonances \cite{Rivera2016}, and perfect UGRs \cite{Yin2020, Lee2020}.

Anisotropic planar waveguides are known to support leaky modes above the light line (see  \cite{Marcuse1979, Knoesen1988, Torner1993, Lu1993}, and references therein). They are intrinsically hybrid, i.e., they comprise the six electromagnetic field components, and arise as complex solutions of the eigenvalue equation where, for moderate leakage losses, the imaginary part of the eigenvalue provides a good approximation of the radiation losses via the radiation channels. It has recently been established that under suitable conditions in structures with a single radiation channel, radiation of the leaky modes can be totally cancelled, thus yielding anisotropy-induced BICs \cite{Gomis-Bresco2017, Mukherjee2018, Mukherjee2019, Pankin2020}. Anisotropy induced BICs have also been used for diverse applications \cite{Yu2019,Yu_2020}. It must be properly appreciated that, due to their full-vector hybrid structure, such states are of a totally different nature than BICs arising in systems described with scalar transverse electric and transverse magnetic mode sets. One salient consequence is the profound role that anisotropy-symmetry (dictated by the orientations of the optical axes of the materials relative to the waveguide geometry and to each other), rather than standard material symmetry, has on the light propagation phenomena that are possible.

In this Letter, we address structures with two radiation channels and show that, under proper conditions, they can  support UGRs where radiation is cancelled in one of the radiation channels and fully redirected into the other. More specifically, we analyze anti-guiding waveguide structures using the Berreman transfer matrix method \cite{Berreman1972}, supplemented by the condition for BIC existence. This formalism readily reveals that the two radiation channels are strongly coupled at the boundaries due to birefringence. We find that UGRs only appear when at least one of the optic axes (OAs) in the structure is taken out of the waveguide plane, breaking the polar anisotropy-symmetry. They are robust and exhibit phase singularities in the radiation channel that is cancelled. Here we focus on describing the physical properties of the UGRs and the practical conditions that make them possible, based on the analysis elaborated in detail in Ref.~\cite{Mukherjee2018}.

\begin{figure}[t]
\centering
\includegraphics[width=0.8\linewidth]{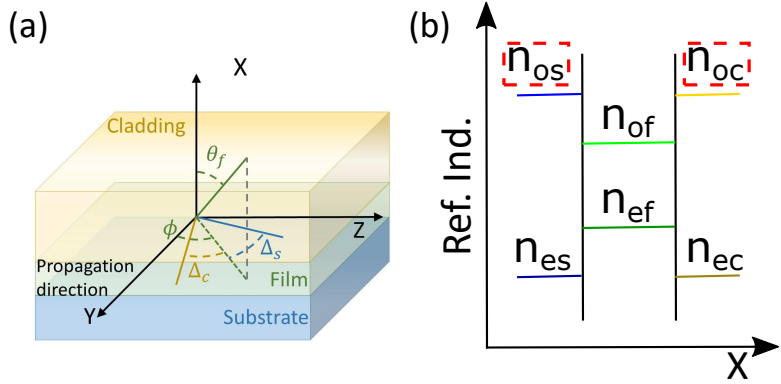}
\caption{(a) Waveguide comprising three negative birefringent materials whose optical axis orientations are assumed to be varied independently. (b) Schematic of the refractive indices of the structure. The dashed red boxes indicate the index/polarisation corresponding to the radiation channel.
\\ \hrulefill}
\label{fig:1}
\end{figure}

\begin{figure}[t]
\centering
\includegraphics[width=0.95\linewidth]{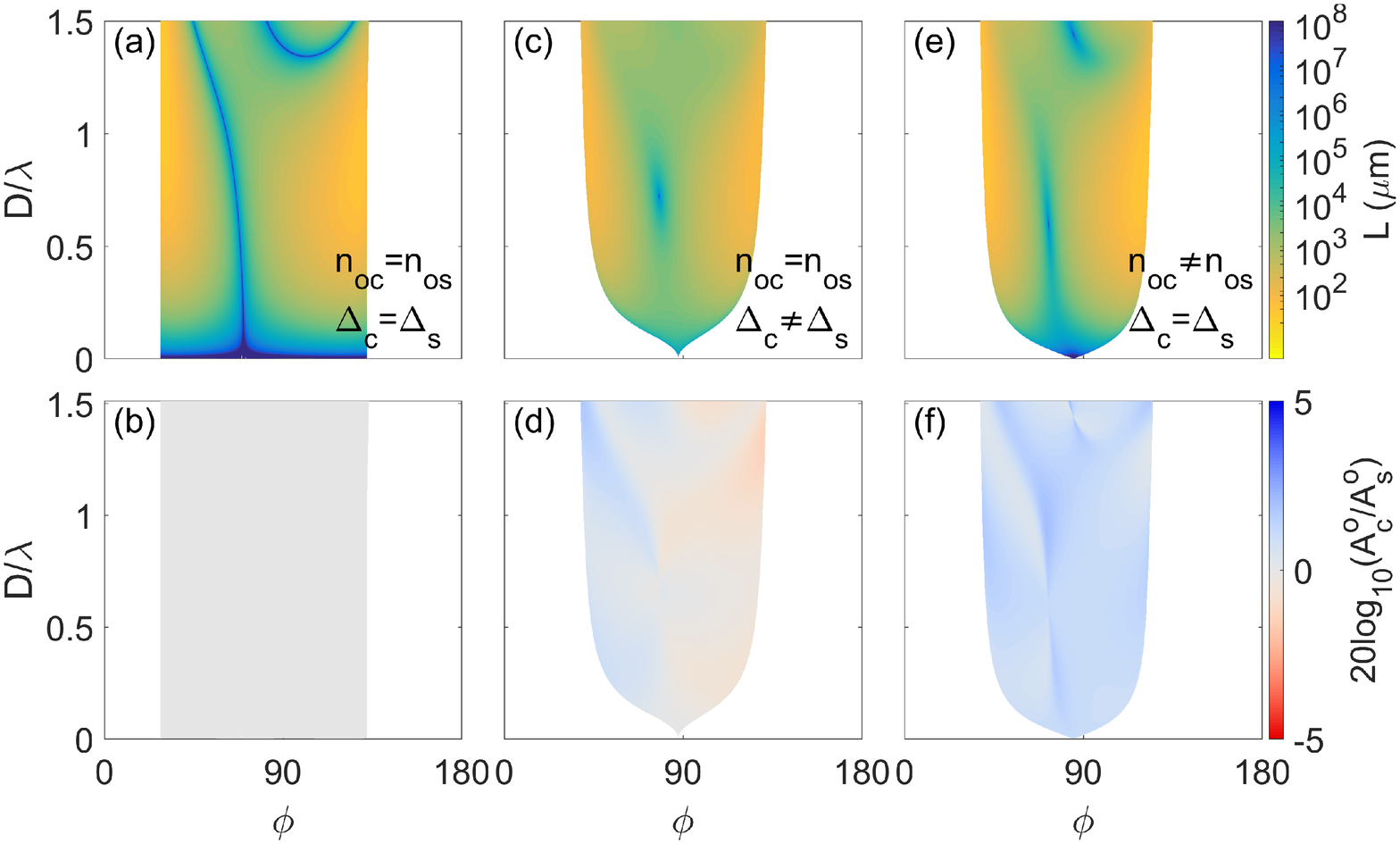}
\caption{(a) Fundamental leaky mode supported by a geometrically symmetric structure with $\Delta_c=\Delta_s=5^{\circ}$, (c) an asymmetric structure with $\Delta_c=0^{\circ}$ and $\Delta_s=5^{\circ}$ and (e) an asymmetric structure with $n_{oc}\neq n_{os}=1.8$. In all cases the OAs are contained in a plane parallel to the interface ($\theta = 90^{\circ}$). The remaining parameters are defined in the main text. The color shows attenuation length $L$ of the leaky mode, defined as the length at which the field amplitude decays to $1/e$ of the initial value. The white area corresponds to areas beyond the mode cutoff. The ratio of radiation channel amplitudes (in dB) is shown in (b, d, f) for the respective structures in (a, c, e).
\\ \hrulefill}
\label{fig:2}
\end{figure}

The class of structures we study is depicted in Fig.~\ref{fig:1}(a). It comprises negative birefringent uniaxial materials in the substrate, the core/film and the cover. $D$ is the thickness of the film and thus $D/\lambda$ is the normalized dimensionless thickness or normalized operating wavelength. We consider propagation along the $y$ direction, while $x$ is orthogonal to all the waveguide interfaces. The angle $\phi$ denotes the azimuthal orientation in the interface plane between the propagation direction and the projection of the film OA. $\Delta_c$ and $\Delta_s$ give the misalignment of the OAs of the cover and the substrate with respect to the film, such that $\Delta_{c/s}=\phi-\phi_{c/s}$. $\theta_s, \theta_f$ and $\theta_c$ are the polar angles of the substrate, film and cover OAs with respect to the normal to the interface plane. Without loss of generality, unless otherwise specified, here we consider identical uniaxial negative materials in the substrate and the cover (i.e., $n_{os}=n_{oc}=1.7>n_{es}=n_{ec}=1.3$), with the refractive indices of the material of the film situated between those of the cover and the substrate ($n_{os}>n_{of}=1.6>n_{ef}=1.4>n_{es}$), as shown in Fig.~\ref{fig:1}(b). The ordinary index of the cover and substrate being the highest refractive index of the structure ensures that no guided modes can be supported and the ordinary polarisation in the substrate and the cover always provide the two aforementioned radiation channels via which light in the film can couple to the radiation continuum. It is assumed that the OAs of the substrate, film and cover are  independently oriented during the fabrication of the waveguide. 

The radiation channels are equivalent when the structure is mirror-symmetric about the $x=0$ plane and it is expected that such structures will support lines of BIC existence in a manner analogous to anisotropic waveguides with a single radiation channel. Figure \ref{fig:2}(a) shows the fundamental semi-leaky mode supports lines of interferometric (INT) BICs (blue lines) for a structure that maintains mirror-symmetry even though azimuthal anisotropy-symmetry is broken ($\Delta_c=\Delta_s\neq 0^{\circ}$). Figure  \ref{fig:2}(b) shows the ratio of the radiation channel amplitudes in dB for the mode in Fig.~\ref{fig:2}(a). The symmetric structure has equivalent radiation channels and therefore the radiation in the two channels is identical. In the case that the mirror symmetry about $x=0$ is broken, radiation into the two channels is asymmetric as shown in Figs.~\ref{fig:2}(c-f) for two different perturbations of the mirror symmetry. Figure \ref{fig:2}(c) shows the fundamental semi-leaky mode in a structure where the OAs in the cover and the substrate are no longer parallel, namely $\Delta_c=0^{\circ} \neq \Delta_s=5^{\circ}$. Then, BIC lines collapse to a BIC point, performing a topological transition in the dispersion diagram \cite{Mukherjee2018}. In addition, the asymmetry in the structure leads to some asymmetry in the radiation into the two channels, as shown in Fig.~\ref{fig:2}(d) but this is not substantial. If the asymmetry in the structure arises from the use of different materials in the cover and the substrate, e.g., with $n_{oc}\neq n_{os}$ while keeping $\Delta_c=\Delta_s$, once again the BIC lines collapse to BIC points [Fig.~\ref{fig:2}(e)] and yet again the radiation in the two channels is asymmetric but not substantially [see Fig.~\ref{fig:2}(e)]. In both cases, complete unidirectional suppression of the radiation is not observed. 

\begin{figure}[t]
\centering
\includegraphics[width=0.95\linewidth]{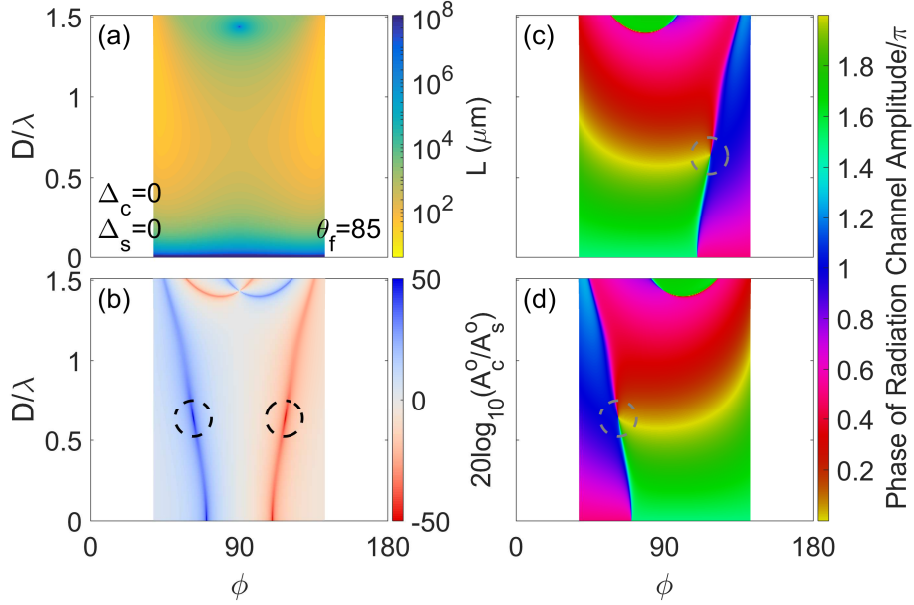}
\caption{(a) Fundamental leaky mode supported by a structure with only polar anisotropy-symmetry breaking ($\Delta_c=\Delta_s=0^{\circ}$ and $\theta_f=85^{\circ}$). (b) Ratio of radiation channel amplitudes in dB for the mode in (a). The dashed black circles in (b) mark the unidirectional guided resonances. (c) Phase of the radiation channel (ordinary wave) amplitude in the cover, measured with respect to the phase of the extraordinary confined wave. (d) Same as (c) but for the substrate radiation channel amplitude. The dashed grey circles in (c,d) show the screw phase dislocations in the radiation channel amplitudes. 
\\ \hrulefill}
\label{fig:3}
\end{figure}

\begin{figure}[t]
\centering
\includegraphics[width=0.95\linewidth]{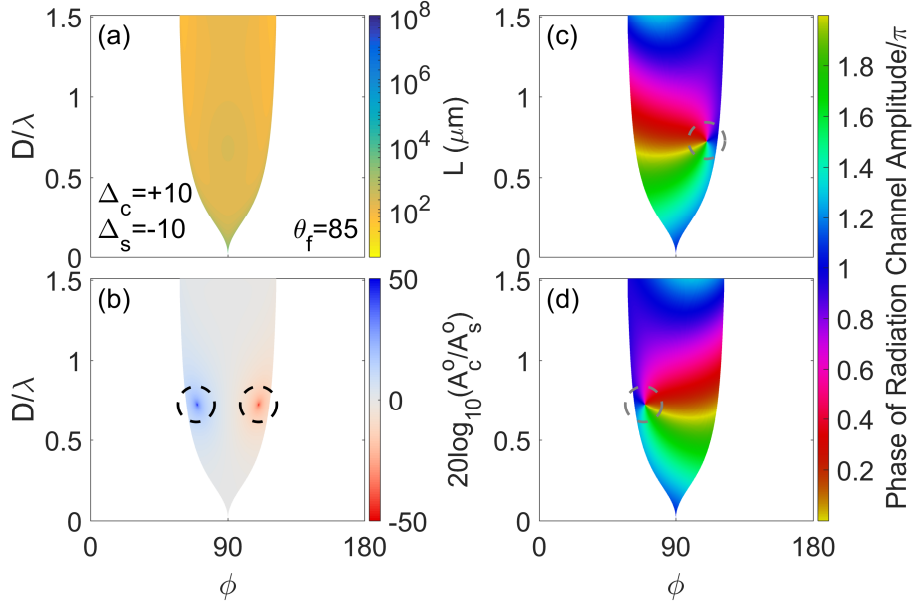}
\caption{Same as Fig.~\ref{fig:3} but for both azimuthal and polar anisotropy-symmetry breaking $\Delta_c=-\Delta_s=10^{\circ}$ and $\theta_f=85^{\circ}$.
\\ \hrulefill}
\label{fig:4}
\end{figure}

The mirror symmetry with respect to the $x=0$ plane can also be broken by taking at least one of the OAs out of the interface plane, i.e., by breaking the polar anisotropy symmetry. Figure \ref{fig:3} shows the fundamental semi-leaky mode when the film OA is at  $\theta_f=85^{\circ}$. Even though $\Delta_c=\Delta_s= 0^{\circ}$, the structure is geometrically asymmetric and therefore the equivalence of the radiation channels is also broken. Again, BIC lines collapse to BIC points, as the one shown at $\phi=90^{\circ}$ and $D/\lambda\approx 1.434$ as shown in Fig.~\ref{fig:3}(a) (more BIC points exist above $D/\lambda > 1.5$). The radiation into the two different channels is distinct as shown in Fig.~\ref{fig:3}(b) but now there is a relevant qualitative difference. The blue and red lines indicate strongly asymmetric radiation (note the change in limits of the color scale) to the cover and the substrate, respectively. Moreover, there are specific points (dashed black circles in Fig.~\ref{fig:3}(b)) in the $\phi-D/\lambda$ space where the ratio diverges. At these points radiation into one channel is entirely canceled while all radiation escapes via the other channel. Thus, such points correspond to unidirectional guided resonances \cite{Yin2020}. Because $\Delta_c=\Delta_s=0$, the two UGRs occur at the same value of $D/\lambda=0.643$, and at symmetric directions about $\phi=90^{\circ}$, in this case specifically at  $\phi=62.09^{\circ}$ and $\phi=117.91$, radiating to the cover and the substrate, respectively. As Fig.~\ref{fig:3}(a) shows, radiation losses do not decrease at the UGR (the color scale is homogeneous at the UGR point). Therefore, the UGR in one radiation channel deviates all the radiation to the opposite channel on the leaky mode.


Topological transitions in the dispersion diagram of the structures from BIC lines to BIC points caused by polar anisotropy-symmetry breaking result in zeroes of radiation characterised by screw phase singularities in the radiation channel amplitude at the BIC point \cite{Mukherjee2018}. This is the case for the BIC point in Fig.~\ref{fig:3}(a), which results in screw phase singularities at exactly the same point, $\phi-D/\lambda$, for the two channels, as shown by the phase of the radiation channel amplitudes in the cover and the substrate, as depicted in Figs.~\ref{fig:3}(c) and (d). The winding number is opposite in the substrate and cover radiation channels, as could be observed by the gradual change in the phase, which is only appreciable upon zooming in the plot. The UGRs correspond to screw phase singularities (grey dashed circles in Figs.~\ref{fig:3}(c,d)) that do not coincide in the $\phi-D/\lambda$ space for the two channels, indicating that radiation is zero only in the corresponding channel, which has become decoupled from the continuum due to destructive interference. The UGRs in this system are characterised by integer winding numbers assigned to their corresponding phase singularities unlike the UGRs in photonic crystals that are assigned half integral charges \cite{Yin2020}.

We also found that it is possible to switch the perfectly unidirectional radiation from the cover to the substrate by changing the polar OA orientation to $\theta_f=95^{\circ}$. Then, the position of the UGRs is interchanged, so that the UGR at $\phi=62.09^{\circ}$ radiates to the substrate and the UGR at $\phi= 117.91^{\circ}$ radiates to the cover. Changing the polar OA orientation is equivalent to a $180^{\circ}$ rotation of the structure, or to reversing the direction of propagation from $+y$ to $-y$. This kind of switching is a characteristic property of the system we are addressing and cannot be realised in a structure with only one radiation channel where one of the cladding materials forbids radiation.

Figure \ref{fig:4} shows the fundamental semi-leaky mode for a structure where both polar and azimuthal anisotropy-symmetry is broken ($\theta_f=85^{\circ}$ and $\Delta_c= -\Delta_s= 10^{\circ}$). In this case, there are no BICs supported on the fundamental semi-leaky mode as shown in Fig.~\ref{fig:4}(a). The lines of strongly asymmetric radiation to one radiation channel present in Fig.~\ref{fig:3}(b) also disappear in Fig.~\ref{fig:4}(b) but the UGRs survive, as shown by the dashed black circles Fig.~\ref{fig:4}(b). Again, the cancellation of radiation in one channel is compensated by an increase of radiation into the other channel, so that the total loses on the leaky mode, Fig.~\ref{fig:4}(a), are not altered by the presence of the UGRs.  Like in the case of Fig. \ref{fig:3}, the UGRs appear at symmetric positions with respect $\phi=90^{\circ}$ and correspond to screw phase singularities in the radiation channel which has been decoupled from continuum [Figs.~\ref{fig:4}(c,d)]. 

We therefore establish that UGRs are possible only under polar anisotropy-symmetry breaking, as azimuthal anisotropy-symmetry breaking or material asymmetry in the structure does not create them. If $\Delta_c=-\Delta_s$, UGRs are located at mirror symmetric positions about $\phi=90^{\circ}$ on the leaky mode. With polar anisotropy symmetry broken and $\Delta_c\neq-\Delta_s$, UGRs continue to exist though they are no longer placed in symmetric positions about $\phi=90^{\circ}$ on the leaky mode. When the UGRs for the two radiation channels coincide in the $\phi-D/\lambda$ space, a BIC is created, as is the case of the BIC point in Fig. ~\ref{fig:3}. Similarly, when $\theta_f$ returns to $\theta_f=90^{\circ}$ and the structure evolves from the situation with broken polar anisotropy-symmetry to the situation where polar anisotropy-symmetry is restored, the two UGRs in Fig.~\ref{fig:3} follow a trajectory in the $\phi-D/\lambda$ space until they coincide at $\phi=90^{\circ}$ and form the BIC lines in Fig.~\ref{fig:2}(a). Therefore, UGRs only appear/disappear when BICs are broken/restored, and as the screw phase singularity indicates, they are robust against any perturbation. 

\begin{figure}[t]
\centering
\includegraphics[width=\linewidth]{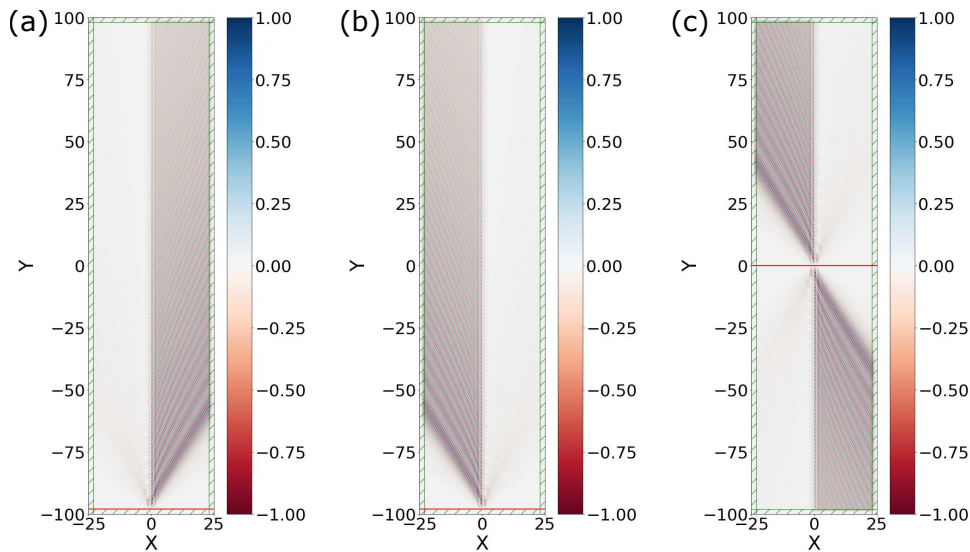}
\caption{FDTD calculation of unidirectional guided resonances, showing the $z$-component of the magnetic field ($H_z$), in a structure with only polar anisotropy symmetry breaking ($\theta_f=85^{\circ}$ and $\Delta_c=\Delta_s= 0^{\circ}$) for  $D/\lambda=0.643$ and (a) $\phi=62.09^{\circ}$ and (b-c) $\phi=117.91^{\circ}$. (a) and (b) corresponds to forward propagation with radiation into the cover and substrate, respectively. (c) Simultaneous forward and backwards propagation. The $Y$ propagation distance is normalized relative to the wavelength. 
\\ \hrulefill}
\label{fig:5}
\end{figure}

The light propagation dynamics that occurs at the UGR is shown by finite-difference time-domain (FDTD) calculations using MEEP \cite{MeepCitation}. The results are shown in Fig.~\ref{fig:5} for the structure  Fig.~\ref{fig:3} with $\theta_f= 85^{\circ}$. The UGR radiating only to the cover with $\phi=62.09^{\circ}$ and $D/\lambda=0.643$ (left circle in Fig.~\ref{fig:3}(b)) is shown in Fig.~\ref{fig:5}(a), resulting in radiation to the cover. Fig.~\ref{fig:5}(b) shows the UGR radiating only to the substrate at $\phi=117.91^{\circ}$.  Fig.~\ref{fig:5}(c) shows the scenario where the structure is excited in the centre with both forward and backward propagation with $\phi=117.91^{\circ}$. For forward propagation, in the $+y$ direction,  the situation does not change, and radiation  goes to the substrate, as in Fig.~\ref{fig:4}(b). A change of propagation direction to $-y$ results in radiation going to the cover, as the situation is equivalent to having $\phi=62.09^{\circ}$ or orienting the film OA at $\theta_f= 95^{\circ}$. Note that in the FDTD calculations, we use the eigenmode source provided by MEEP, which  does not exactly match the improper leaky mode, resulting in some reshaping during propagation.

Therefore, we conclude that unidirectional guided resonances can exist in anisotropic anti-guiding waveguides with multiple radiation channels only when polar anisotropy-symmetry is broken. When the polar anisotropy-symmetry is not broken, the UGRs in the two channels coincide to form a BIC. Thus, the UGRs arise from the same mechanism as anisotropy induced BICs. Anisotropy-induced UGRs show screw phase singularities and their existence conditions are robust under variation of system parameters. They also allow for switching of the perfectly unidirectional radiation from cover to substrate and vice versa.  These findings highlight the new phenomena introduced by the concept of (natural or artificial) anisotropy-symmetry and are relevant to the practical applications of UGRs in waveguide structures made of birefringent materials whose optical axis orientation can be varied during the fabrication process or during operation, such as liquid crystals. Unidirectional radiation may have important applications in several optoelectronic devices such as on-chip lasers, directional optical antennas or directional couplers \cite{Yin2020}, and our results open the possibility to implement them in off-axis geometries in anisotropic materials.


{\bf Funding}. H2020 Marie Sklodowska-Curie Action GA665884; Government of Spain (grants PGC2018-097035-B-I00; Severo Ochoa CEX2019-000910-S); Fundació Cellex; Fundació Mir-Puig; Generalitat de Catalunya (CERCA and AGAUR 2017-SGR-1400).

{\bf Disclosures}. The authors declare no conflicts of interest.

\bibliography{biblio}

\begin{thebibliography}{10}

\bibitem{Yin2020}
X.~Yin, J.~Jin, M.~Soljačić, C.~Peng, and B.~Zhen, ``Observation of
  topologically enabled unidirectional guided resonances,'' {\em Nature},
  vol.~580, pp.~467--471, Apr. 2020.

\bibitem{Hirose2014}
K.~Hirose, Y.~Liang, Y.~Kurosaka, A.~Watanabe, T.~Sugiyama, and S.~Noda,
  ``Watt-class high-power, high-beam-quality photonic-crystal lasers,'' {\em
  Nature Photonics}, vol.~8, pp.~406--411, May 2014.

\bibitem{Taillaert2004}
D.~Taillaert, P.~Bienstman, and R.~Baets, ``Compact efficient broadband grating
  coupler for silicon-on-insulator waveguides,'' {\em Optics Letters}, vol.~29,
  p.~2749, Dec. 2004.

\bibitem{Vermeulen2010}
D.~Vermeulen, S.~Selvaraja, P.~Verheyen, G.~Lepage, W.~Bogaerts, P.~Absil,
  D.~Van~Thourhout, and G.~Roelkens, ``High-efficiency fiber-to-chip grating
  couplers realized using an advanced {CMOS}-compatible
  {Silicon}-{On}-{Insulator} platform,'' {\em Optics Express}, vol.~18,
  p.~18278, Aug. 2010.

\bibitem{Yaacobi2014}
A.~Yaacobi, J.~Sun, M.~Moresco, G.~Leake, D.~Coolbaugh, and M.~R. Watts,
  ``Integrated phased array for wide-angle beam steering,'' {\em Optics
  Letters}, vol.~39, p.~4575, Aug. 2014.

\bibitem{Roncone1993}
R.~L. Roncone, L.~Li, K.~A. Bates, J.~J. Burke, L.~Weisenbach, and B.~J.~J.
  Zelinski, ``Design and fabrication of a single leakage-channel grating
  coupler,'' {\em Applied Optics}, vol.~32, p.~4522, Aug. 1993.

\bibitem{Kim2006}
S.-H. Kim, S.-K. Kim, and Y.-H. Lee, ``Vertical beaming of wavelength-scale
  photonic crystal resonators,'' {\em Physical Review B}, vol.~73, June 2006.

\bibitem{Wade2015}
M.~T. Wade, F.~Pavanello, R.~Kumar, C.~M. Gentry, A.~Atabaki, R.~Ram,
  V.~Stojanović, and M.~A. Popović, ``75\% efficient wide bandwidth grating
  couplers in a 45 nm microelectronics {CMOS} process,'' in {\em 2015 {IEEE}
  {Optical} {Interconnects} {Conference} ({OI})}, pp.~46--47, Apr. 2015.
\newblock ISSN: 2376-8665.

\bibitem{Wang2013}
K.~X. Wang, Z.~Yu, S.~Sandhu, and S.~Fan, ``Fundamental bounds on decay rates
  in asymmetric single-mode optical resonators,'' {\em Optics Letters},
  vol.~38, p.~100, Jan. 2013.

\bibitem{Ota2015}
Y.~Ota, S.~Iwamoto, and Y.~Arakawa, ``Asymmetric out-of-plane power
  distribution in a two-dimensional photonic crystal nanocavity,'' {\em Optics
  Letters}, vol.~40, p.~3372, July 2015.

\bibitem{Zhou2016}
H.~Zhou, B.~Zhen, C.~W. Hsu, O.~D. Miller, S.~G. Johnson, J.~D. Joannopoulos,
  and M.~Soljačić, ``Perfect single-sided radiation and absorption without
  mirrors,'' {\em Optica}, vol.~3, p.~1079, Oct. 2016.

\bibitem{Neuman1929}
J.~von Neuman and E.~Wigner, ``Uber merkwürdige diskrete {Eigenwerte}. {Uber}
  das {Verhalten} von {Eigenwerten} bei adiabatischen {Prozessen},'' {\em Z.
  Phys.}, vol.~30, pp.~467--470, 1929.

\bibitem{Stillinger1975}
F.~H. Stillinger and D.~R. Herrick, ``Bound states in the continuum,'' {\em
  Phys. Rev. A}, vol.~11, pp.~446--454, Feb. 1975.

\bibitem{Marinica2008}
D.~C. Marinica, A.~G. Borisov, and S.~V. Shabanov, ``Bound states in the
  continuum in photonics,'' {\em Phys. Rev. Lett.}, vol.~100, p.~183902, May
  2008.

\bibitem{Bulgakov2008}
E.~N. Bulgakov and A.~F. Sadreev, ``Bound states in the continuum in photonic
  waveguides inspired by defects,'' {\em Phys. Rev. B}, vol.~78, p.~075105, Aug
  2008.

\bibitem{Plotnik2011}
Y.~Plotnik, O.~Peleg, F.~Dreisow, M.~Heinrich, S.~Nolte, A.~Szameit, and
  M.~Segev, ``Experimental observation of optical bound states in the
  continuum,'' {\em Phys. Rev. Lett.}, vol.~107, p.~183901, Oct 2011.

\bibitem{Corrielli2013}
G.~Corrielli, G.~Della~Valle, A.~Crespi, R.~Osellame, and S.~Longhi,
  ``Observation of surface states with algebraic localization,'' {\em Phys.
  Rev. Lett.}, vol.~111, p.~220403, Nov 2013.

\bibitem{Hsu2013}
C.~W. Hsu, B.~Zhen, J.~Lee, S.-L. Chua, S.~G. Johnson, J.~D. Joannopoulos, and
  M.~Soljačić, ``Observation of trapped light within the radiation
  continuum,'' {\em Nature}, vol.~499, pp.~188--191, July 2013.

\bibitem{Gomis-Bresco2017}
J.~Gomis-Bresco, D.~Artigas, and L.~Torner, ``Anisotropy-induced photonic bound
  states in the continuum,'' {\em Nat. Photon.}, vol.~11, pp.~232--236, Mar.
  2017.

\bibitem{Monticone2017}
F.~Monticone and A.~Alù, ``Bound states within the radiation continuum in
  diffraction gratings and the role of leaky modes,'' {\em New J. Phys.},
  vol.~19, no.~9, p.~093011, 2017.

\bibitem{Minkov2018}
M.~Minkov, I.~A.~D. Williamson, M.~Xiao, and S.~Fan, ``Zero-index bound states
  in the continuum,'' {\em Phys. Rev. Lett.}, vol.~121, p.~263901, Dec 2018.

\bibitem{Azzam2018}
S.~I. Azzam, V.~M. Shalaev, A.~Boltasseva, and A.~V. Kildishev, ``Formation of
  {Bound} {States} in the {Continuum} in {Hybrid} {Plasmonic}-{Photonic}
  {Systems},'' {\em Physical Review Letters}, vol.~121, Dec. 2018.

\bibitem{Fan2019}
K.~Fan, I.~V. Shadrivov, and W.~J. Padilla, ``Dynamic bound states in the
  continuum,'' {\em Optica}, vol.~6, pp.~169--173, Feb 2019.

\bibitem{Liang2020}
Y.~Liang, K.~Koshelev, F.~Zhang, H.~Lin, S.~Lin, J.~Wu, B.~Jia, and Y.~Kivshar,
  ``Bound {States} in the {Continuum} in {Anisotropic} {Plasmonic}
  {Metasurfaces},'' {\em Nano Letters}, June 2020.

\bibitem{Kodigala2017}
A.~Kodigala, T.~Lepetit, Q.~Gu, B.~Bahari, Y.~Fainman, and B.~Kanté, ``Lasing
  action from photonic bound states in continuum,'' {\em Nature}, vol.~541,
  pp.~196--199, Jan. 2017.

\bibitem{Romano2018}
S.~Romano, A.~Lamberti, M.~Masullo, E.~Penzo, S.~Cabrini, I.~Rendina, and
  V.~Mocella, ``Optical {Biosensors} {Based} on {Photonic} {Crystals}
  {Supporting} {Bound} {States} in the {Continuum},'' {\em Materials}, vol.~11,
  p.~526, Mar. 2018.

\bibitem{Carletti2018}
L.~Carletti, K.~Koshelev, C.~De~Angelis, and Y.~Kivshar, ``Giant nonlinear
  response at the nanoscale driven by bound states in the continuum,'' {\em
  Phys. Rev. Lett.}, vol.~121, p.~033903, Jul 2018.

\bibitem{Hayran2021}
Z.~Hayran and F.~Monticone, ``Capturing broadband light in a compact bound
  state in the continuum,'' {\em ACS Photonics}, 2021.

\bibitem{Hsu2016}
C.~W. Hsu, B.~Zhen, A.~D. Stone, J.~D. Joannopoulos, and M.~Soljačić, ``Bound
  states in the continuum,'' {\em Nat. Rev. Mat.}, vol.~1, Sept. 2016.

\bibitem{Zhen2014}
B.~Zhen, C.~W. Hsu, L.~Lu, A.~D. Stone, and M.~Solja\ifmmode \check{c}\else
  \v{c}\fi{}i\ifmmode~\acute{c}\else \'{c}\fi{}, ``Topological nature of
  optical bound states in the continuum,'' {\em Phys. Rev. Lett.}, vol.~113,
  p.~257401, Dec 2014.

\bibitem{Bulgakov2017}
E.~N. Bulgakov and D.~N. Maksimov, ``Topological bound states in the continuum
  in arrays of dielectric spheres,'' {\em Phys. Rev. Lett.}, vol.~118,
  p.~267401, Jun 2017.

\bibitem{Doeleman2018}
H.~M. Doeleman, F.~Monticone, W.~den Hollander, A.~Alù, and A.~F. Koenderink,
  ``Experimental observation of a polarization vortex at an optical bound state
  in the continuum,'' {\em Nature Photonics}, vol.~12, pp.~397--401, July 2018.

\bibitem{Zhang2018}
Y.~Zhang, A.~Chen, W.~Liu, C.~W. Hsu, B.~Wang, F.~Guan, X.~Liu, L.~Shi, L.~Lu,
  and J.~Zi, ``Observation of polarization vortices in momentum space,'' {\em
  Phys. Rev. Lett.}, vol.~120, p.~186103, May 2018.

\bibitem{Mukherjee2018}
S.~Mukherjee, J.~Gomis-Bresco, P.~Pujol-Closa, D.~Artigas, and L.~Torner,
  ``Topological properties of bound states in the continuum in geometries with
  broken anisotropy symmetry,'' {\em Phys. Rev. A}, vol.~98, p.~063826, Dec
  2018.

\bibitem{Jin2019}
J.~Jin, X.~Yin, L.~Ni, M.~Soljačić, B.~Zhen, and C.~Peng, ``Topologically
  enabled ultrahigh-{Q} guided resonances robust to out-of-plane scattering,''
  {\em Nature}, vol.~574, pp.~501--504, Oct. 2019.

\bibitem{Rivera2016}
N.~Rivera, C.~W. Hsu, B.~Zhen, H.~Buljan, J.~D. Joannopoulos, and
  M.~Soljačić, ``Controlling {Directionality} and {Dimensionality} of
  {Radiation} by {Perturbing} {Separable} {Bound} {States} in the
  {Continuum},'' {\em Sci. Rep.}, vol.~6, p.~33394, Dec. 2016.

\bibitem{Lee2020}
S.-G. Lee, S.-H. Kim, and C.-S. Kee, ``Bound states in the continuum (bic)
  accompanied by avoided crossings in leaky-mode photonic lattices,'' {\em
  Nanophotonics}, no.~0, p.~20200346, 03 Aug. 2020.

\bibitem{Marcuse1979}
D.~Marcuse and I.~Kaminow, ``Modes of a symmetric slab optical waveguide in
  birefringent media - {Part} {II}: {Slab} with coplanar optical axis,'' {\em
  IEEE J. Quantum Electron.}, vol.~15, pp.~92--101, Feb. 1979.

\bibitem{Knoesen1988}
A.~Knoesen, T.~K. Gaylord, and M.~G. Moharam, ``Hybrid guided modes in uniaxial
  dielectric planar waveguides,'' {\em J. Lightw. Technol.}, vol.~6,
  pp.~1083--1104, Jun 1988.

\bibitem{Torner1993}
L.~Torner, J.~Recolons, and J.~P. Torres, ``Guided-to-leaky mode transition in
  uniaxial optical slab waveguides,'' {\em J. Lightw. Technol.}, vol.~11,
  pp.~1592--1600, Oct. 1993.

\bibitem{Lu1993}
M.~Lu and M.~M. Fejer, ``Anisotropic dielectric waveguides,'' {\em Journal of
  the Optical Society of America A}, vol.~10, p.~246, Feb. 1993.

\bibitem{Mukherjee2019}
S.~Mukherjee, J.~Gomis-Bresco, P.~Pujol-Closa, D.~Artigas, and L.~Torner,
  ``Angular control of anisotropy-induced bound states in the continuum,'' {\em
  Optics Letters}, vol.~44, p.~5362, Nov. 2019.

\bibitem{Pankin2020}
P.~S. Pankin, B.-R. Wu, J.-H. Yang, K.-P. Chen, I.~V. Timofeev, and A.~F.
  Sadreev, ``One-dimensional photonic bound states in the continuum,'' {\em
  Communications Physics}, vol.~3, Dec. 2020.

\bibitem{Yu2019}
Z.~Yu, X.~Xi, J.~Ma, H.~K. Tsang, C.-L. Zou, and X.~Sun, ``Photonic integrated
  circuits with bound states in the continuum,'' {\em Optica}, vol.~6,
  pp.~1342--1348, Oct 2019.

\bibitem{Yu_2020}
Z.~Yu, Y.~Tong, H.~K. Tsang, and X.~Sun, ``High-dimensional communication on
  etchless lithium niobate platform with photonic bound states in the
  continuum,'' {\em Nature Communications}, vol.~11, Dec. 2020.

\bibitem{Berreman1972}
D.~W. Berreman, ``Optics in {Stratified} and {Anisotropic} {Media}:
  4x4-{Matrix} {Formulation},'' {\em J. Opt. Soc. Am.}, vol.~62, pp.~502--510,
  Apr. 1972.

\bibitem{MeepCitation}
A.~F. Oskooi, D.~Roundy, M.~Ibanescu, P.~Bermel, J.~Joannopoulos, and S.~G.
  Johnson, ``Meep: {A} flexible free-software package for electromagnetic
  simulations by the {FDTD} method,'' {\em Computer Physics Communications},
  vol.~181, pp.~687--702, Mar. 2010.

\end{thebibliography}
\bibliographystyle{ieeetr}

\end{document}